\documentclass[]{spie}  

\usepackage{makecell} 
\usepackage{booktabs} 
\usepackage{array}
\usepackage{stackengine}
\usepackage{amsmath,amsfonts,amssymb}
\usepackage{graphicx}
\usepackage[colorlinks=true, allcolors=blue]{hyperref}

\title{Comparison of Phase-Unwrapping Methods for Adaptive Optics Wavefront Sensing}

\author[a]{D. Angelica Huerta}
\author[a]{Justin R. Crepp}
\author[a]{Caleb G. Abbott}
\author[a]{Brian Joseph}

\affil[a]{University of Notre Dame, 225 Nieuwland Science Hall, Notre Dame, IN 46556 USA}

\authorinfo{Further author information: (Send correspondence to D. Angelica Huerta)\\D. Angelica Huerta: E-mail: dhuerta2@nd.edu}

\begin{document} 
\maketitle

\begin{abstract}
Advanced wavefront sensors (WFS) are essential for enabling new science cases for telescopes that utilize adaptive optics (AO) systems. While complex field WFS --- those that estimate the electric field phase and amplitude through interference or diffraction effects --- can achieve extraordinary sensitivity compared to existing devices, they typically reconstruct the wrapped phase of the measured wavefront, which must then be unwrapped for correction by continuous-surface deformable mirrors (DM). Another requirement is that the phase function must be unwrapped within 1 millisecond or faster for real-time AO operations. Using simulations of atmospheric turbulence that follow a Kolmogorov spectrum, we study four prevalent and mature phase unwrapping methods: Fast2D, Zernike Gradient (Zernike), Discrete Fourier Transform (DFT), and Least Squares Principle Value (LSPV). In this paper, we examine the strengths and limitations of each method. In particular, we consider performance with and without a binary circular aperture boundary that defines the edge of monolithic telescopes. 
\end{abstract}


\section{INTRODUCTION}\label{sec:intro}

\begin{figure}[ht]\label{fig:F1}
    \centering
        \includegraphics[width=\textwidth]{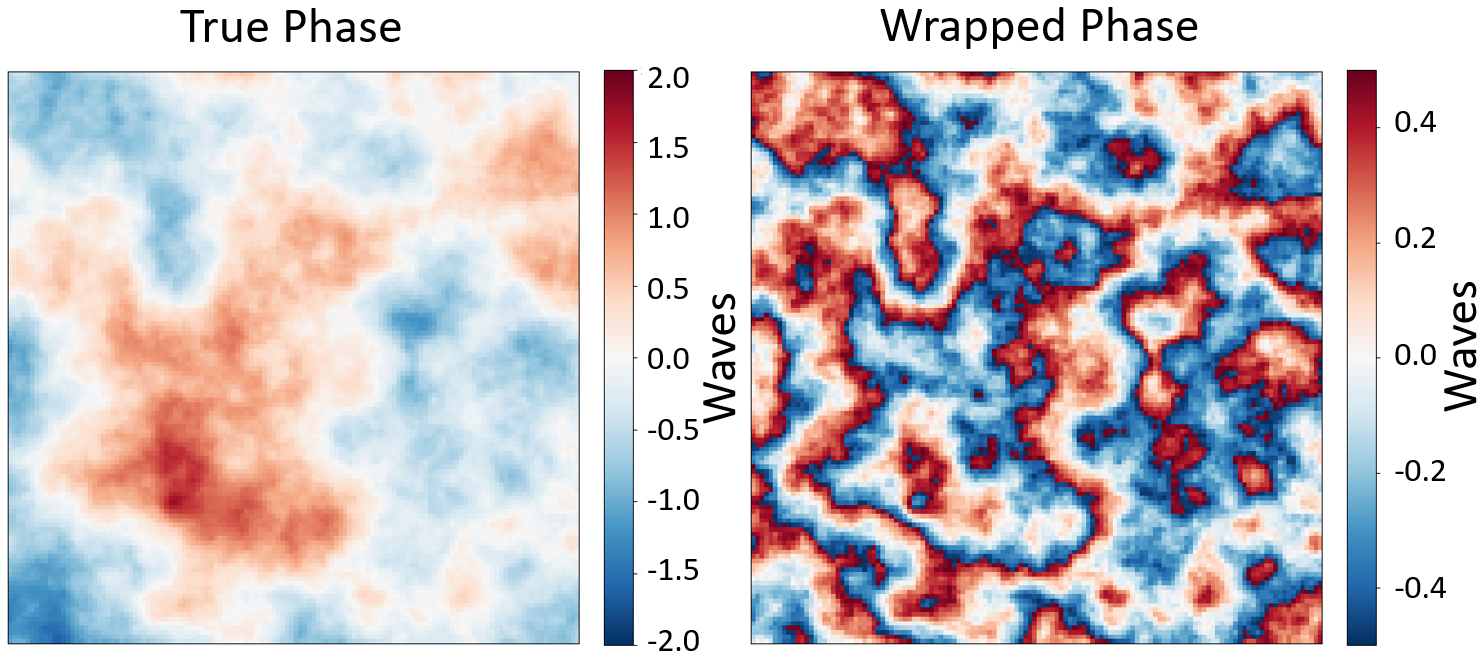} 
     \caption{Example simulated Kolmogorov wavefront (left) and wrapped phase (right).}
\end{figure}

Wavefront sensors (WFS) are an integral component of adaptive optics (AO) systems, used in various applications such as high-resolution imaging and spectroscopy\cite{macintosh2014,crepp2014}, remote sensing\cite{Miyamura2011}, power beaming\cite{leland1992}, and laser communications\cite{Tyson1996}. For this reason, advancing WFS technologies is essential for improving the imaging quality delivered by AO systems. Many WFS return the wrapped phase ($\Phi_w$) --- the modulo-2$\pi$ representation of the true phase ($\Phi$) --- following reconstruction due to degeneracies in the electric field that result from measured periodic interference fringes. The phase wrapping phenomena occurs when the reconstructed phase ``wraps around" a $2 \pi$ interval (Figure~\ref{fig:F1}), usually confined to [$0$ to 2$\pi$) or ($-\pi$ to $\pi$] based on the inversion of trigonometric functions. As a consequence, the WFS output includes spatially abrupt discontinuities known as phase wraps~\cite{ghiglia1998,pellizzari2010}, 
\begin{equation}\label{phase_wrapping}
\Phi_w = \mathrm{Arg}(e^{i\Phi}) = \tan^{-1}\left(\frac{\mathrm{Im}(e^{i\Phi})}{\mathrm{Re}(e^{i\Phi})}\right) = \Phi - 2\pi \; \mathrm{floor}\left[\frac{\Phi + \pi}{2\pi}\right] \nonumber \\
\end{equation}
where the ``floor'' operator rounds down to the nearest integer. These discontinuities introduce inaccuracies in phase measurements, particularly in applications like optical interferometry and remote sensing, where phase wrapping obscures true phase information. For instance, in interferometry, phase wrapping can lead to erroneous fringe patterns~\cite{huntley1995}, reducing the accuracy of measurements and image reconstruction. Ultimately, the discontinuities caused by phase wrapping complicate data analysis and interpretation, making it necessary to recover the true phase from the wrapped measurements through a process known as phase unwrapping. Since most AO systems employ a deformable mirror (DM) with a continuous face-sheet, it is necessary to unwrap the phase using algorithms that reconstitute the wavefront. 

Additionally, the WFS dictates the ultimate speed of the AO system. Thus, it is necessary to unwrap the phase both accurately (spatial) and efficiently (temporal).\cite{tyson2010} To keep pace with high-speed wavefront acquisition, unwrapping must be done within a millisecond to accommodate changes in atmospheric turbulence. Recent advances in computational power, involving CPU's and parallelization, GPU's, FPGA's, firmware, and low-latency data transfer, make it feasible to consider more sophisticated unwrapping methods that rapidly reconstruct wavefront phases.  

Numerous phase unwrapping algorithms have been developed to address the phase wrapping problem, each offering unique approaches to overcoming this challenge. These algorithms are typically organized into the following five categories\cite{pellizzari2010,yu2019}: 
\begin{enumerate}
    \item Path-Following: Among the most well-studied unwrapping methods, path-following algorithms adjust phase values along a chosen path to minimize the risk of errors by either starting from a seed point and expanding outwards or by using a quality map to guide the unwrapping process\cite{goldstein1988,flynn1997}. 
    
    \item Regional: These methods, like path-following algorithms, typically follow a minimum-risk path, but instead of unwrapping the entire phase at once, they divide the phase map into regions that are unwrapped independently and subsequently merged \cite{ghiglia1994}.
    
    \item Global: By treating phase unwrapping as an optimization problem, which is resolved by solving a system of equations that minimizes phase inconsistencies and discontinuities across the entire phase map, global algorithms ensure the best overall solution for the phase. Globally distributed error corrections are computationally intensive\cite{pritt1994}.

    \item AI/ML: Artificial Intelligence \& Machine Learning (AI/ML) unwrapping algorithms use machine learning models, such as Convolutional Neural Networks (CNN's), to learn from data and directly infer the true phase from the wrapped phase by automatically identifying and correcting phase ambiguities\cite{Spoorthi2019}. 

    \item Hybrid: Hybrid methods combine multiple phase unwrapping techniques to form a more effective approach. By integrating different strategies, they enhance either efficiency or accuracy compared to the original methods.
\end{enumerate}
The goal of this paper is to evaluate and compare several phase unwrapping algorithms to identify those capable of meeting the stringent performance demands of modern AO. While unwrapping is a well-studied problem, achieving both speed and accuracy, in real-time remains challenging. We focus on methods that show potential for sub-millisecond execution while maintaining robustness under varying boundary constraints. Section~\ref{sec:algos} provides a brief overview of four algorithms (Fast2D, Zernike, DFT, and LSPV) that show promise for further enhancement. Section~\ref{sec:analysis} details the analysis and comparisons of these algorithms under different conditions. Finally, Section~\ref{sec:Conclusion} concludes the paper with a discussion of the results and future research directions.
\section{ALGORITHMS EXPLORED}\label{sec:algos}

The algorithms explored in this study represent a non-exhaustive list that were chosen because they demonstrate promising preliminary results in terms of accuracy and efficiency, are well-suited for AO, and offer potential for further improvement. In this section, the methodology of each algorithm is described, noting benefits and potential drawbacks as identified in the literature and through experience.

\subsection{Fast2D Method}\label{sec:fast2d Method}

The Fast2D phase unwrapping algorithm\cite{Herráez2002} is a quality-guided, path-following algorithm that prioritizes resolving the most reliable edges first, following a non-continuous path. The method first calculates the second difference ($D$) of every pixel, to determine pixel reliability ($R = 1/D$), following the approach described in Herráez et al. 2002\cite{Herráez2002}. Subsequently, the edge reliability is calculated by summing the reliability of neighboring phase values. The edges are then sorted according to their reliability, and the algorithm unwraps phase values in this order. The fundamental premise of this unwrapping method is that adjacent phase values that are similar are likely to constitute regions where phase wrapping has not occurred and should therefore be unwrapped first.

An advantage of the Fast2D algorithm is its use of simple arithmetic to calculate pixel reliability and unwrap the phase, making it both straightforward and highly accurate. Readily available implementations in Matlab\cite{Firman2025}, Python, and C++ have also been extensively tested by researchers, as evidenced by the large number of citations \cite{Herráez2002}. One limitation of the algorithm is that edge pixels lack neighboring data to compute the second difference, requiring care around edges or sharp boundaries, such as with optical systems whose beam may be defined by a circular telescope, central obstruction, and support bars. This algorithm feature also places limitations on the ability to perform spatial parallelization, since adjacent segments must overlap and require unwrapping iterations to stitch together results from individual regions.  

\subsection{Zernike Gradient Method}\label{sec:zern Method}

The Zernike Gradient phase unwrapping algorithm (hereafter called ``Zernike'') is a global algorithm proposed by Guyon\cite{guyon2010}. The algorithm relies on the fact that phase slopes remain constant across wrapped boundaries. The process begins by calculating the gradient of the wrapped phase in the $x$ and $y$ directions. Zernike polynomial coefficients are then reconstructed from the $x$ and $y$ gradients using least-squares fitting. Coefficients of the wavefront are unwrapped by reconstructing the phase as a sum of the Zernike polynomials using gradient measurements. In other words, a predetermined set of Zernike polynomials are used in a linear combination to best fit the wavefront slopes and reconstruct the unwrapped wavefront.

The Zernike algorithm is conceptually easy to understand, operates efficiently, and is highly effective for managing circular telescope geometries since Zernike polynomials are expressed in polar coordinates. Additionally, the number of Zernike polynomials may be adjusted, allowing for a trade-off between accuracy and latency. Lastly, implicit regularization by using spatial modes may offer benefits in noisy environments. 

\subsection{Discrete Fast Fourier Transform Method}\label{sec:DFT Method}

The Discrete Fast Fourier Transform (hereafter called ``DFT'') algorithm by Schofield and Zhu\cite{Schofield2003} is a global algorithm based on the principle that the derivative of the phase remains continuous across phase wraps. By extension, the second derivative should also be continuous, or at least well-behaved. Schofield and Zhu’s key insight was that Fourier transforms may be used to relate the wrapped phase to the unwrapped phase in closed-form using Laplacians. 

The DFT unwrapper is a computationally efficient algorithm that relies on Fast Fourier transforms (FFT) and, in principle, requires only several FFT operations to perform phase unwrapping. However, several challenges limit its performance. First, reconstruction accuracy degrades near boundary edges due to discontinuities and rounding errors in the integer number of waves to unwrap, $n(r)$. Correcting the rounding errors requires an iterative process, slowing down the algorithm. Second, the use of an inverse Laplacian introduces unwanted periodic boundary conditions into the solution for $n(r)$. To mitigate this issue, Schofield \& Zhu recommend using cosine transforms or a larger array size of $2N\times2N$ to enforce symmetry, albeit at the expense of computational complexity and latency. Despite these limitations, the DFT unwrapper is considered a promising method due to its mathematical elegance. 

\subsection{Least Squares Principal Value Method}\label{sec:LSPV Method}

The basic premise of the Least Squares Principal Value (hereafter called ``LSPV'') unwrapping method is to shear the phase map with itself (in both the $x$ and $y$ directions) and study the resulting difference in phase values, similar to taking a spatial derivative. The algorithm solves a least-squares reconstruction problem using discrete phase difference operators to constrain phase jumps to the principal $[-\pi,\pi]$ range. LSPV is a global algorithm first proposed by Barchers\cite{barchers2002}, with multiple variations existing, including the original version implemented in the WaveProp Matlab program\cite{Lata2025}. Later developments include the LSPV+N series of unwrappers, which incorporate the Postprocessing Congruence Operation (PCO), and culminate with the LSPV+7 unwrapper\cite{Steinbock2014}. Recent work further shows that such methods are still being worked with because they are essential when hidden-phase components dominate, as in strong turbulence or with extended beacons\cite{Kalensky2025}. 

One of the main advantages of the LSPV algorithm is its adaptability—multiple implementations exist, allowing users to choose between higher accuracy or faster computation. Preliminary tests and evaluations by teams in industry indicate that LSPV outperforms its predecessors, the $S$-phase and $X$-phase unwrappers\cite{Steinbock2014}, in both accuracy and computational speed.   
\section{ANALYSIS AND COMPARISONS}\label{sec:analysis}

To compare performance of the phase-unwrapping algorithms described in 
Section~\ref{sec:algos}, we evaluated their accuracy and latency under controlled conditions using Monte Carlo simulations. Accuracy is quantified by calculating root-mean-square (RMS) wavefront error (WFE, measured in waves) between the piston-removed unwrapped phase and known true phase, while latency is estimated in milliseconds using a CPU profiler. The analysis was conducted in two stages. First, we examine idealized square wavefronts by simulating $N=300$ noiseless but turbulent $(D/r_0 = 20)$, $128\times128$ pixel Kolmogorov phase screens, following tip and tilt removal\cite{Abbott2025,Abbott2025b}. While not representative of typical optical system data, this analysis establishes a baseline for comparing algorithm behavior under different circumstances. Then, we introduce circular apertures prior to unwrapping (simulating telescope pupils), to reassess performance and quantify degradation caused by binary boundary conditions. In the former case (unbounded wavefronts), although the full wavefront is unwrapped, results are calculated in post-processing using a circular mask to ensure direct comparison of RMS WFE between scenarios. The mask is undersized by 5\% in diameter to help minimize edge artifacts from unwrapping. 

\subsection{Spatial Analysis}\label{sec:spatial}
\begin{figure}[ht]
    \centering
\includegraphics[height=17cm]{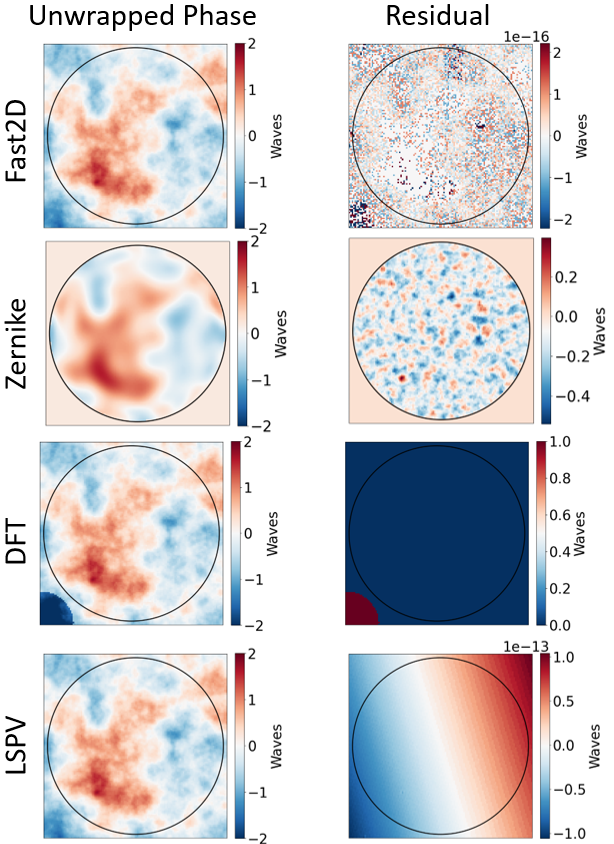}
\caption{Accuracy comparison between phase unwrapping algorithms using the Kolmogorov wavefront from Figure~\ref{fig:F1}.} 
\label{fig:F2} 
\end{figure}

\begin{figure}[ht]
\begin{center}
\includegraphics[height=18cm]{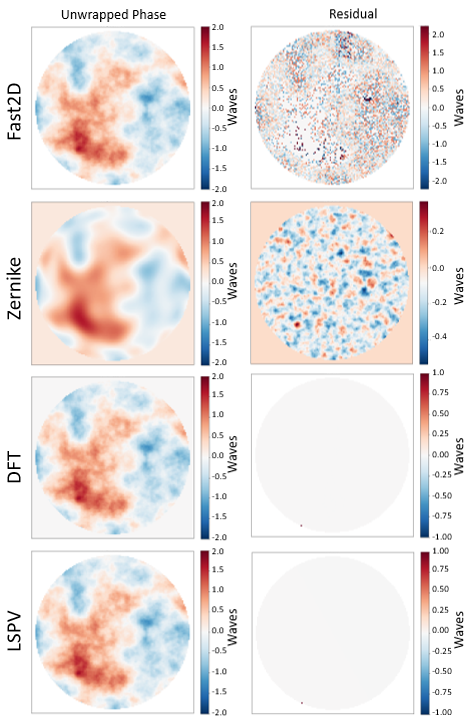}
\end{center}
     \caption{Example phase unwrapping results for the Kolmogorov wavefront shown in Fig.~\ref{fig:F1} when using a circular telescope boundary.}
     \label{fig:F5} 
\end{figure}

As a representative example, Figure~\ref{fig:F2} shows results for the Fast2D, Zernike, DFT, and LSPV phase unwrappers using the wavefront from Figure~\ref{fig:F1}, with a mask applied in postprocessing, while Figure~\ref{fig:F5} shows the results for the same wavefront including aperture boundaries. Each method produces a unique residuals signature depending on the underlying algorithm. After performing simulations that average the results for a statistically significant sample of Kolmogorov wavefronts ($N=300$ trials), patterns began to emerge regarding overall performance and reliability. Results are tabulated in Table 1. 

\begin{table}[ht]
\label{tab:algorithms}
\centering
\begin{tabular}{c|cccc} 
\hline
\hline
Unwrapper & Fast2D & Zernike  & DFT & \makecell{LSPV} \\
\hline
Reference & Herraez (2002)\cite{Herráez2002}  & Guyon (2010)\cite{guyon2010} & Schofield \& Zhu (2003) \cite{Schofield2003} & Barchers (2002)\cite{barchers2002} \\
Language & Python & Python & Matlab & Matlab \\
WFE (no bndry) & $(6.80 \pm 0.04) \times10^{-17} $ & $0.152\pm 0.003$ & $0.33\pm 0.01$ & $(5.1 \pm 0.2) \times10^{-9}$  \\
WFE (w/ bndry) & $(1.15 \pm 0.03) \times10^{-16}$ & $0.153\pm0.003$  & $0.18\pm 0.01$ & $0.065 \pm 0.002$\\
Latency [ms] & $3.01 \pm 0.01$ & $9.067 \pm 0.003$  & $7.94 \pm 0.06$ & $3.93 \pm 0.04$ \\
\hline
\hline
\end{tabular}
\caption{Comparison of phase unwrapping algorithms using a $128 \times 128$ array. WFE is measured in waves RMS.} 
\end{table}

\begin{figure}[ht]
\begin{center} 
\includegraphics[height=8.75cm]{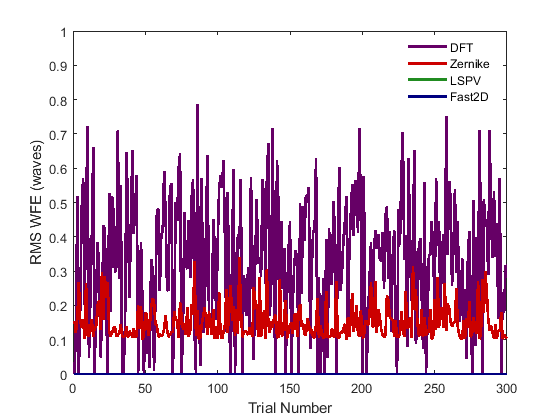} \\
\includegraphics[height=8.75cm]{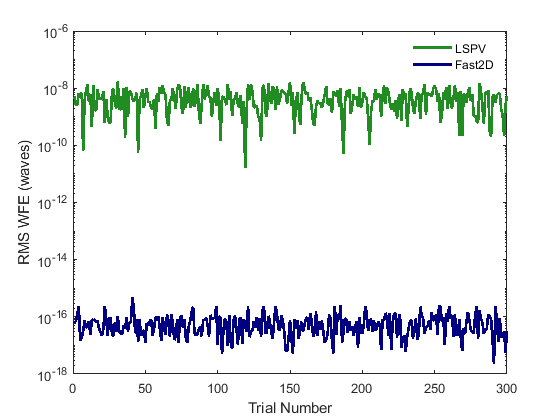}
\end{center}
\caption{(Top:) RMS WFE for 300 wavefront trials without circular telescope boundaries. (Bottom:) Zoomed in view of the LSPV and Fast2D results.}
\label{fig:rms square}
\end{figure}

In the absence of a circular telescope boundary, we find that the Fast2D and LSPV unwrappers routinely reach their respective numerical noise floors, achieving near-perfect accuracy (negligible WFE) without photon noise or camera read noise. Fast2D and LSPV results are repeated on a separate (logarithmic) scale for viewing purposes (Fig.~\ref{fig:rms square}). The Zernike unwrapper reaches diffraction-limited performance, limited only by fitting errors based on the number of modes (200 Zernike polynomials were used for the analysis). 
 
\begin{figure}[ht]
\begin{center}
\includegraphics[height=8cm]{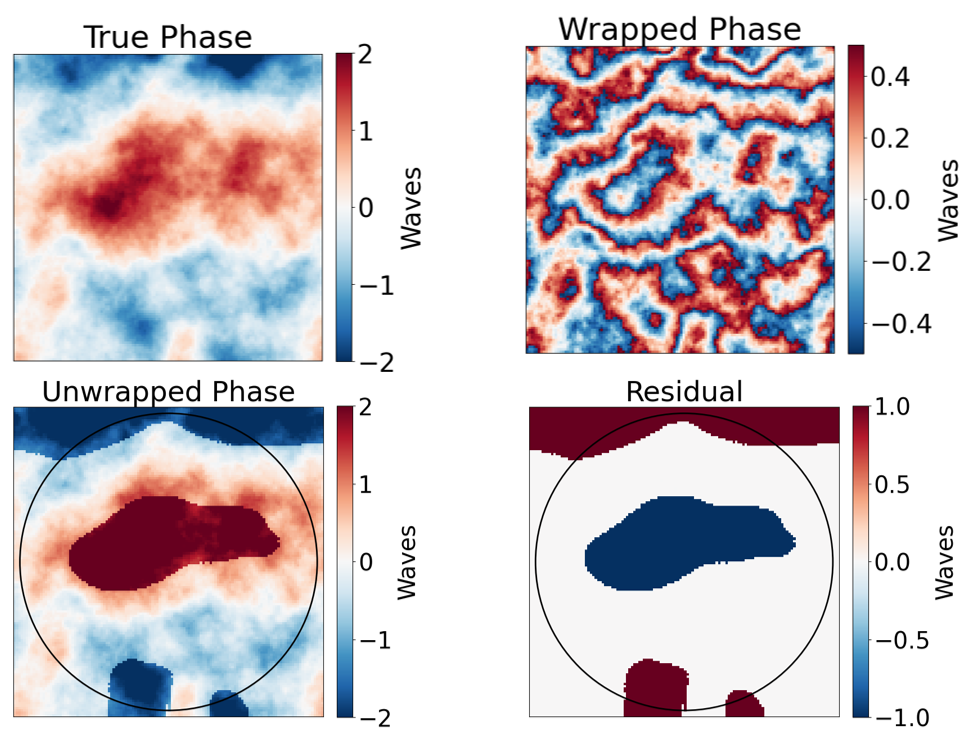}
\end{center}
     \caption{Example unwrapping results for the DFT algorithm. Localized patches of the wavefront are offset by $\pm$ 1-2 waves. This type of behavior is found to occur primarily where the phase varies rapidly.} 
     \label{fig:F3} 
\end{figure}

Meanwhile, the DFT unwrapper shows less consistent results and does not reach diffraction-limited performance, despite allowing for up to 100 iterations. While DFT successfully unwrapped some wavefronts (Figure~\ref{fig:F2}), its average RMS WFE of $0.33$ waves reveals fundamental limitations. Figure~\ref{fig:F3} illustrates how localized patches appear elevated or depressed by a uniform pedestal relative to neighboring phase values. Empirically, we find that these systematic errors, which are visually apparent in the residuals map, tend to be located in regions of the wavefront that experience extreme phase variations. 

In the presence of a circular telescope boundary, we find that the Fast2D algorithm continues to offer outstanding performance as it is only affected in the outer-most edge pixels, which we have masked, regardless of pupil shape in principle. While the LSPV unwrapper experiences measurable degradation with the inclusion of an aperture boundary, it still reaches diffraction-limited performance. Meanwhile, the accuracy of the Zernike unwrapper remains consistent because the polynomial basis set is defined in polar coordinates.   

We find that (perhaps ironically) the DFT unwrapper performs better with aperture boundaries, likely because masking removes edge discontinuities that would otherwise disproportionately influence results in the region of interest and distort its solution. By masking edges, the unwrapper focuses on optimizing uniformity within the region of interest, avoiding artifacts caused by external discontinuities.

\subsection{Latency Analysis}\label{sec:latency}

In addition to accuracy, we evaluated the computational performance of each unwrapping method by measuring latency with the same wavefronts used above. In terms of absolute timing, none of the algorithms were able to reach the 1 millisecond level for $128 \times 128$ arrays using a CPU. In terms of relative timing, we find that Fast2D and LSPV are the most efficient algorithms, each being a factor of several times faster than the DFT or Zernike methods (Table 1). While implementation differences across different high-level programming languages limits definitive conclusions that can be drawn about speed, our experience is that the Fast2D and LSPV are intrinsically faster than DFT and Zernike.\footnote{A non-iterative version of the DFT algorithm was only $\approx 10\%$ faster.}. Whether these methods can be implemented into a real-time AO system depends on how amenable the algorithms are to parallelization and firmware development.   

\begin{figure}[ht]
\begin{center}  
\includegraphics[height=8cm]{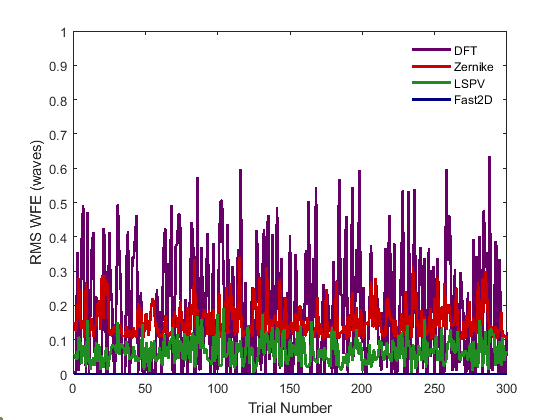} \\ 
\includegraphics[height=8cm]{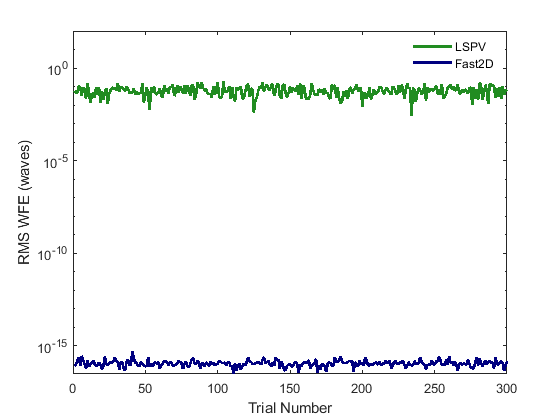}
\end{center}
\caption{(Top:) RMS WFE for phase unwrapping trials with circular telescope boundaries. (Bottom:) Zoomed in view of the LSPV and Fast2D results.}
\label{fig:rms}
\end{figure}

\begin{figure}[ht]
\begin{center}
\includegraphics[height=8cm]{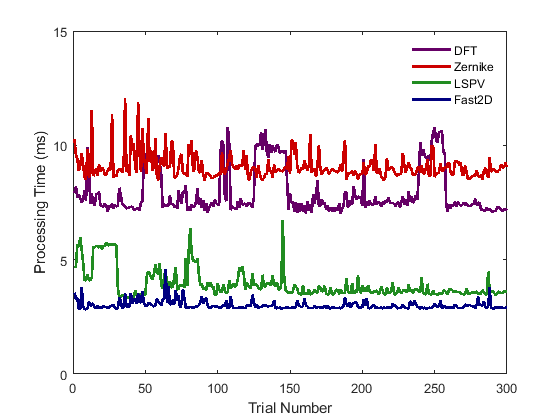} 
\end{center}
\caption{Unwrapping latency for wavefronts with circular boundaries.}\label{fig:latency}
\end{figure} 
\section{CONCLUSION}\label{sec:Conclusion}

In this study, we evaluated four phase unwrapping algorithms (Fast2D, Zernike, DFT and LSPV) for the purpose of AO wavefront sensing considering different boundary conditions. Results are compiled in Table 1, from which we draw the following conclusions: 
\begin{itemize}
    \item The Fast 2D algorithm is the most accurate and lowest-latency method and represents an excellent choice for phase unwrapping. In the case of real-time AO, it is advisable to explore parallelization methods that quantify any losses in unwrapping accuracy due to the increased presence of edges that result from breaking up the wavefront into numerous spatial segments and stitching together solutions.

    \item The LSPV unwrapper ranked second among studied unwrappers in terms of accuracy and latency, offering near-perfect performance without boundaries and diffraction-limited performance with aperture boundaries. For real-time applications, the most time-consuming operation (taking a pseudo-inverse) presents computational challenges for parallelization (see however Lipitakis et al. 2020\cite{LIPITAKIS2020}). Hardware accelerated versions of the LSPV algorithm warrant further exploration.  
    
    \item The Zernike gradient unwrapper also delivered diffraction-limited performance and was impervious to the presence of circular aperture boundaries (as expected). Relative to other algorithms, parallelization of the Zernike gradient algorithm is comparatively straight-forward. Accelerating and optimizing the Zernike unwrapper will be presented in a forthcoming article (Huerta et al. 2026, in prep.).

    \item The DFT algorithm does not consistently achieve diffraction-limited performance, is sensitive to the presence of circular aperture boundaries, and has the highest latency. Although mitigation methods for handling circular boundaries exist, i.e. involving geometric folding by copying the aperture to enforce symmetry (or through the use of cosine transforms), these approaches further sacrifice speed. Although elegant in its simplicity, the DFT algorithm may not be competitive for real-time applications in its current form. 
\end{itemize}

Future work will focus on testing unwrapping methods using more complicated pupil geometries (central obstructions and secondary mirror support structures), different strengths of turbulence, and the inclusion of photon noise and camera read noise. Likewise, we will explore opportunities for latency reduction with algorithm parallelization and hardware acceleration. Such studies are necessary to enable complex-field WFS to meet the stringent demands of next-generation AO systems.

\section*{CODE AND DATA AVAILABILITY}
\label{sec:Availability}
Programs used to generate figures for this article are available upon request from the authors.

\section*{DISCLOSURES}
\label{sec:disclosures}
The authors declare there are no financial interactions, commercial affiliations, or other potential conflicts of interest that have influenced the objectivity of this research of the writing of this article.  

\acknowledgments 
This work was supported in part by the Joint Directed Energy Transition Office (JDETO) and Air Force Office of Scientific Research (AFOSR) grant number FA9550-22-1-0435.

\bibliography{report} 
\bibliographystyle{spiebib} 

\end{document}